\documentclass[journal]{IEEEtran}

\usepackage{amsmath}
\usepackage{graphicx}
\usepackage{color}
\usepackage{cite}

\begin{document}
	
	
	\graphicspath{{images/}}

	\title{3-D Super-Resolution Ultrasound (SR-US) Imaging \\ with a 2-D Sparse Array}
	
	\author{\IEEEauthorblockN{Sevan~Harput, Kirsten Christensen-Jeffries, Alessandro Ramalli, Jemma Brown, Jiaqi Zhu, Ge Zhang,\\ Chee Hau Leow, Matthieu Toulemonde, Enrico Boni, Piero Tortoli, \\ Robert J. Eckersley$^{*}$, Chris Dunsby$^{*}$, and Meng-Xing~Tang$^{*}$}
		
		\thanks{S. Harput, J. Zhu, G. Zhang, C. H. Leow, M. Toulemonde, and M. X. Tang are with the ULIS Group, Department of Bioengineering, Imperial College London, London, SW7 2AZ, UK.}
		\thanks{K. Christensen-Jeffries, J. Brown, and R. J. Eckersley are with the Biomedical Engineering Department, Division of Imaging Sciences, King's College London, SE1 7EH, London, UK. }
		\thanks{A. Ramalli is with the Department of Information Engineering, University of Florence, 50139 Florence, IT and Lab. on Cardiovascular Imaging \& Dynamics, Dept. of Cardiovascular Sciences, KU Leuven, Leuven, Belgium.}
		\thanks{E. Boni, P. Tortoli are with the Department of Information Engineering, University of Florence, 50139 Florence, IT.}
		\thanks{C. Dunsby is with the Department of Physics and the Centre for Pathology, Imperial College London, London, SW7 2AZ, UK.}
		\thanks{$^*$These authors contributed equally to this work.}
		\thanks{E-mail: s.harput@imperial.ac.uk and mengxing.tang@imperial.ac.uk }
	}
	
	\maketitle

	\begin{abstract}
		\boldmath
High frame rate 3-D ultrasound imaging technology combined with super-resolution processing method can visualize 3-D microvascular structures by overcoming the diffraction limited resolution in every spatial direction. However, 3-D super-resolution ultrasound imaging using a full 2-D array requires a system with large number of independent channels, the design of which might be impractical due to the high cost, complexity, and volume of data produced.

In this study, a 2-D sparse array was designed and fabricated with 512 elements chosen from a density-tapered 2-D spiral layout. High frame rate volumetric imaging was performed using two synchronized ULA-OP 256 research scanners. Volumetric images were constructed by coherently compounding 9-angle plane waves acquired in 3 milliseconds at a pulse repetition frequency of 3000 Hz. To allow microbubbles sufficient time to move between consequent compounded volumetric frames, a 7-millisecond delay was introduced after each volume acquisition. This reduced the effective volume acquisition speed to 100 Hz and the total acquired data size by 3.3-fold. Localization-based 3-D super-resolution images of two touching sub-wavelength tubes were generated from 6000 volumes acquired in 60 seconds. In conclusion, this work demonstrates the feasibility of 3D super-resolution imaging and super-resolved velocity mapping using a customized 2D sparse array transducer.

	\end{abstract}
	
	\maketitle

	\section{Introduction}

Visualization of the microvasculature beyond the diffraction limited resolution has been achieved by localizing spatially isolated microbubbles through multiple frames. In the absence of tissue and probe motion, localization precision determines the maximum achievable resolution, which can be on the order of several micrometers at clinical ultrasound frequencies~\cite{Viessmann2013,Desailly2015}. If motion is present and subsequently corrected post-acquisition, then the motion correction accuracy can limit the achievable spatial resolution~\cite{Harput2017a,Harput2018}. Researchers demonstrated the use of 2-D super-resolution ultrasound (SR-US) imaging in many different controlled experiments and pre-clinical studies using microbubbles~\cite{Christensen-Jeffries2015,Ackermann2016,Bar-Zion2017,Foiret2017,Harput2017b,Song2018,Opacic2018,Ilovitsh2018} and nanodroplets~\cite{Luke2016,Yoon2018,Zhang2018,Zhang2018a}. These studies generated super-resolved images of 3-D structures using 1-D ultrasound arrays where super-resolution cannot be achieved in the elevational direction. In addition to this, out-of-plane motion cannot be compensated for when data is only acquired in 2-D. However, with the implementation of 3-D SR-US imaging using a 2-D array, diffraction limited resolution can be overcome in every direction and there is then the potential for 3-D motion tracking and correction.

Many studies have contributed to the development of SR-US imaging methods by improving the localization precision, reducing the acquisition time, increasing microbubble tracking accuracy, and extending the super-resolution into the third dimension. These developments are explained in detail by a recent review~\cite{Couture2018}. Researchers mainly employed two different approaches to generate a super-resolution image of a volume by mechanically scanning the volume with a linear probe and stacking 2-D SR-US images, or by using arrays that can acquire volumetric information electronically. Errico \textit{et al.} have taken steps towards 3-D with a coronal scan of an entire rat brain by using 128 elements of a custom-built linear array at a frequency of 15 MHz. Motion of the probe was controlled with a micro-step motor to generate 2-D super-resolution images over different imaging planes at a frame rate of 500 Hz~\cite{Errico2015}. Lin \textit{et al.} performed a 3-D mechanical scan of a rat FSA tumor using a linear array mounted on a motorized precision motion stage synchronized with the imaging system. They generated 3-D super-resolution images by calculating the maximum intensity projection from all 2-D super-resolution slices, acquired using plane-wave imaging with a frame rate of 500 Hz~\cite{Lin2017}. Although sub-diffraction imaging has not been published using a 2-D imaging probe with a high volumetric imaging rate, 3-D super-resolution has been achieved by previous studies. O'Reilly and Hynynen used a subset of 128 elements from a 1372-element hemispherical transcranial therapy array at a rate of 10 Hz. They generated 3-D super-resolution images of a spiral tube phantom through an \textit{ex vivo} human skullcap at an imaging center frequency of 612 kHz~\cite{Reilly2013}. Desailly \textit{et al.} implemented a plane wave ultrafast imaging method using an ultrasound clinical scanner with 128 fully programmable emission-reception channels. They placed 2 parallel series of 64 transducers to image microfluidic channels and obtained 3-D super-localization by fitting parallel parabolas in the elevation direction~\cite{Desailly2013}.  Christensen-Jeffries \textit{et al.} generated volumetric 3-D super-resolution at the overlapping imaging region of two orthogonal transducers at the focus. They used two identical linear arrays to image sub-diffraction cellulose tubes using amplitude modulated plane-wave transmission at 3 MHz with a frame rate of 400 Hz~\cite{Christensen-Jeffries2017a}.

The development of high-speed programmable ultrasound systems and 2-D arrays created new opportunities for volumetric imaging with high spatio-temporal resolution. In parallel to these hardware developments, novel 3-D imaging methods based on small numbers of transmit-receive pairs enabled a more reliable visualization of tissue volumes~\cite{Provost2014}, the analysis of fast and complex blood flow in 3-D~\cite{Pihl2014,Provost2015,Correia2016,Holbek2017}, the characterization of mechanical properties of tissue by 4-D shear-wave imaging~\cite{Provost2014,Gennisson2015}, the tracking of the pulse wave propagation along the arterial wall~\cite{Apostolakis2017}, the estimation of 4-D tissue motion~\cite{Salles2015}, and other \textit{in vivo} transient events. These technological advances in 3-D imaging also offer new opportunities for SR-US. Although volumetric imaging methods have already shown significant benefits for various ultrasound imaging applications, 3-D imaging with large 2-D arrays requires a high number of hardware channels and huge computational power.

\begin{figure}[!t]
	\centering
	\includegraphics[viewport = 30 35 540 548,  width = 80mm, clip]{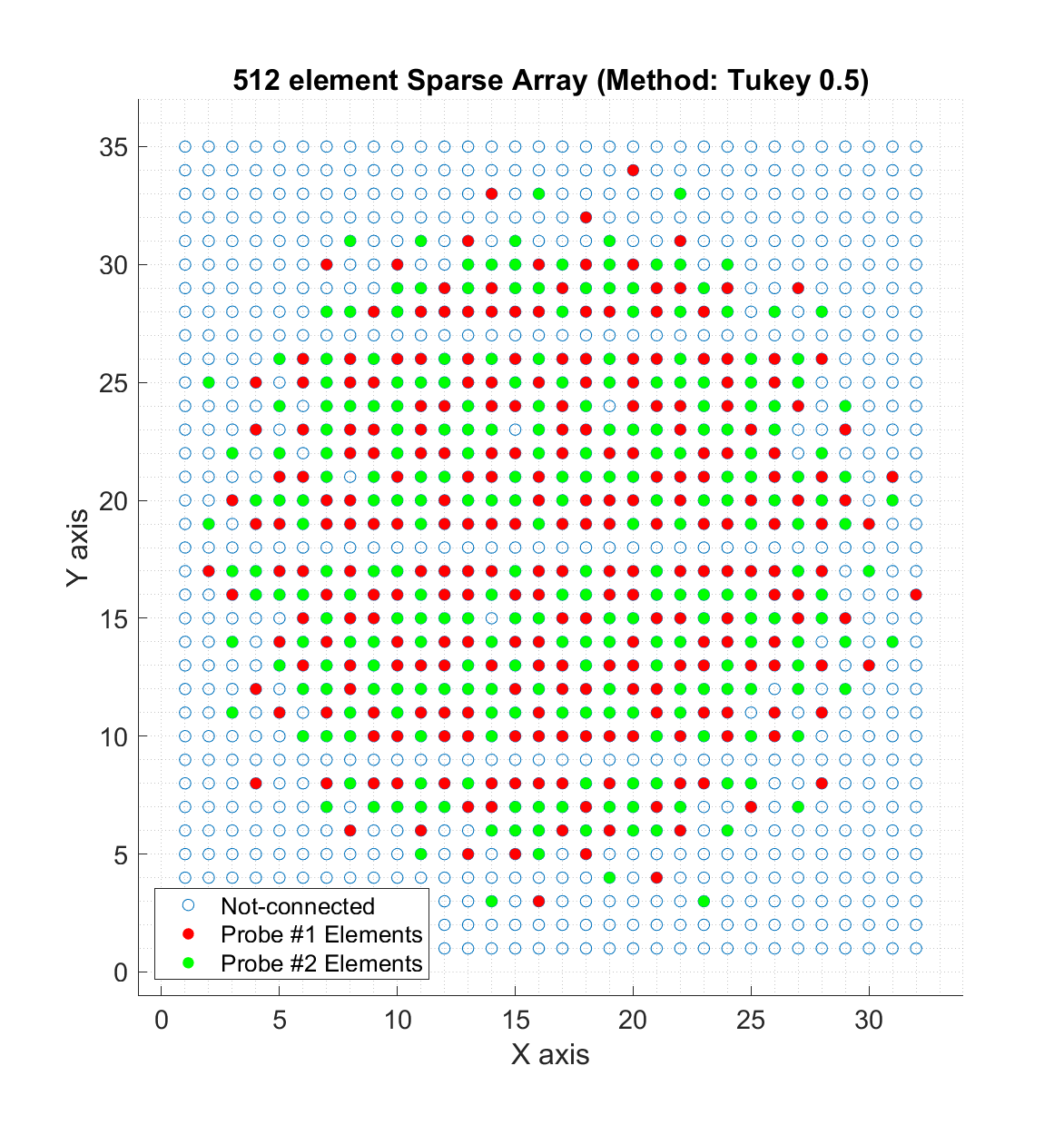}
	\caption{Layout of the 2-D sparse array with red and green circles showing the chosen elements. The pitch between consecutive elements in the $x$ and $y$ directions is 300~$\mu$m. Empty rows (9, 18, and 27) are due to manufacturing limitations and are not related to the density-tapered 2-D spiral method.}
	\label{fig:SparseArray}
\end{figure}

In this study, we demonstrate the feasibility of 3D super-resolution imaging and super-resolved flow velocity mapping using a density-tapered sparse array instead of a full 2-D array to reduce the number of channels and hence the amount of data while maintaining the frame rate. A similar approach was in previous non-super-resolution studies on minimally redundant 2-D arrays~\cite{Karaman2009} and sparse 2-D arrays~\cite{Austeng2002,Diarra2013,Roux2016,Roux2017,Roux2018}, but uses a greater number of elements to improve transmit power and receive sensitivity. Our method significantly differs from row-column addressing and multiplexing approaches since it maintains simultaneous access to all probe elements through independent channels. The sparse array was designed specifically for high volumetric rate 3-D super-resolution ultrasound imaging based on a density-tapered spiral layout~\cite{Ramalli2015a,Harput2018a}. The capability of the 2-D sparse array for 3-D SR-US imaging was demonstrated in simulations and experiments.

	\section{Materials and Methods}
	
	\subsection{2-D Sparse Array}

A 2-D sparse array was designed by selecting 512 elements from a $32 \times 35$ gridded layout of a 2D matrix array (Vermon S.A., Tours, France) as shown in Fig.~\ref{fig:SparseArray}. It was fabricated with an individual element size of $300 \times 300$ $\mu$m, center frequency of 3.7 MHz and a bandwidth of 60\%. In the y direction, row numbers 9, 18, and 27 were intentionally left blank for wiring, hence the total number of available elements is 1024. The method to select the location of sparse array elements is based on the density-tapered 2-D spiral layout~\cite{Ramalli2015a}. This method arranges the elements according to seeds generated from Fermat’s spiral function with an additional spatial density modulation to reduce the side lobes of the transmitted beam profile. This deterministic, aperiodic, and balanced positioning procedure guarantees uniform performance over a wide range of imaging angles.

\begin{figure}[!t]
	\centering
	\includegraphics[width = 88mm]{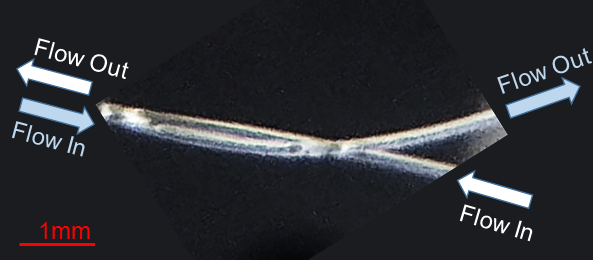}
	\caption{Optical image of two 200 $\mu$m cellulose tubes arranged in a double helix pattern. To create this pattern, two tubes were wrapped around each other which created contact-points that are visible in the optical image. Both tubes had constant microbubble flow in opposite directions. }
	\label{fig:Optical_image}
\end{figure}

It is not possible to connect all 512 elements to a single ultrasound probe adapter. Therefore, two sparse array layouts, hereinafter referred to as Aperture{\#}1 and Aperture{\#}2, were designed as shown with red and green elements in Fig.~\ref{fig:SparseArray}. Both sparse arrays were based on an ungridded, 10.4-mm-wide spiral with 256 seeds~\cite{Ramalli2015a}, whose density tapering was modulated according to a 50\%-Tukey window. The elements belonging to Aperture{\#}1 were selected among those of the Vermon 2-D matrix array, by activating the available elements whose positions were closest to the ideal positions of the ungridded spiral. Similarly, the elements belonging to Aperture{\#}2 were also selected among those of the Vermon matrix array, but excluding those that were already assigned to Aperture{\#}1. The two layouts were connected to two independent connectors (model DLP 408, ITT Cannon, CA, USA) so that an approximation of a 256-element density tapered spiral array could be driven by an independent ULA-OP 256 system~\cite{Boni2016,Boni2017}. Moreover, by synchronizing two ULA-OP 256 systems to simultaneously control the two layouts, a 512-element dense array (Aperture{\#}1 + Aperture{\#}2) with integrated Tukey apodization could be driven. 

	\subsection{Experimental Setup}

Two ULA-OP 256~\cite{Boni2016,Boni2017} systems were synchronized to transmit 9 plane waves from the 512 selected elements. Plane waves were steered within a range of $\pm10^{\circ}$ degrees with a step size of $5^{\circ}$ in the lateral and elevational directions. A 3-cycle Gaussian pulse with a 3.7~MHz center frequency was used for imaging. Pre-beamforming raw data for 9 angles were acquired in 3 milliseconds with a pulse repetition frequency (PRF) of 3000 Hz. These 9 volumetric acquisitions were compounded to construct a singe imaging volume. 
	
The microvessel phantom was made of two 200$\pm$15~$\mu$m Hemophan cellulose tubes (Membrana, 3M, Germany) with a wall thickness of 8$\pm$1~$\mu$m. Two tubes were arranged in a double helix shape  at a depth of 25~mm as shown in Fig.~\ref{fig:Optical_image}. The volumetric B-mode imaging was performed without microbubble flow inside these tubes. For SR-US imaging, a 1:800 diluted Sonovue (Bracco S.p.A, Milan, Italy) solution was flowed through both tubes in opposite directions with a constant flow rate that produced an average microbubble velocity of 10~mm/s using an infusion pump. The PRF of 3000 Hz was high enough to limit motion artefacts on 9 compounded volumes due to moving microbubbles in flow~\cite{Toulemonde2017}. An MI of 0.07 was used for imaging that destroyed a significant number of microbubbles before reaching the end of the tube at this insonation rate. To improve the microbubble longevity, a 7~milliseconds pause was introduced between each compounded volume acquisition that effectively reduced the volumetric acquisition rate to 100 Hz. During this interval, microbubbles travelled approximately 100~$\mu$m, which is less than the size of the B-mode PSF. This short interval between acquisitions also reduced the data size by a factor of 3.3 and maximized the total acquisition duration for the allocated memory size. A total of 6000 volumetric ultrasound frames were acquired in 60 seconds.

	\subsection{Super-resolution Processing and Velocity Calculations}
The RF signals obtained by each aperture ({\#}1 and {\#}2) were separately beamformed. First, singular value decomposition was performed on these datasets to separate the microbubble signal and the echoes from the tube~\cite{Demene2015}. After isolating the microbubble signals, data acquired from two probes were combined offline using the ASAP method~\cite{Stanziola2018}. By processing and beamforming the data from two apertures separately an additional noise reduction step was introduced since a noisy signal resembling a microbubble echo is unlikely to occur simultaneously on both beamformed volumes from different systems.

After combining the beamformed data from both apertures to reconstruct a single volume, an intensity threshold was applied to further reduce the noise level by removing the data below the threshold value. After thresholding, super-localization was performed on the remaining data that may represent a microbubble~\cite{Christensen-Jeffries2017}. In addition to detecting their locations, the size of every microbubble echo was calculated. To remove the localizations that may belong to multiple-microbubbles, detections were discarded if their volume is two times larger than the volume of a pre-calibrated PSF.

Velocities of detected microbubbles were traced using the nearest neighbor method between consecutive frames. Two additional measures were used to filter incorrect pairings. First, if, in consecutive frames, there was more than 70\% deviation in volume size between the microbubble echoes, that velocity track was replaced with the next closest microbubble pair after the same size comparison. Second, the velocity tracks with a value larger than 20~mm/s were discarded, since the maximum possible microbubble velocity is expected to be two times larger than the average microbubble velocity of 10~mm/s for laminar flow.

	\section{Results}

	\subsection{2-D Sparse Array Simulation Results}


\begin{figure}[tb]
	\centering
	\includegraphics[viewport = 70 50 805 560,  width = 88mm, clip]{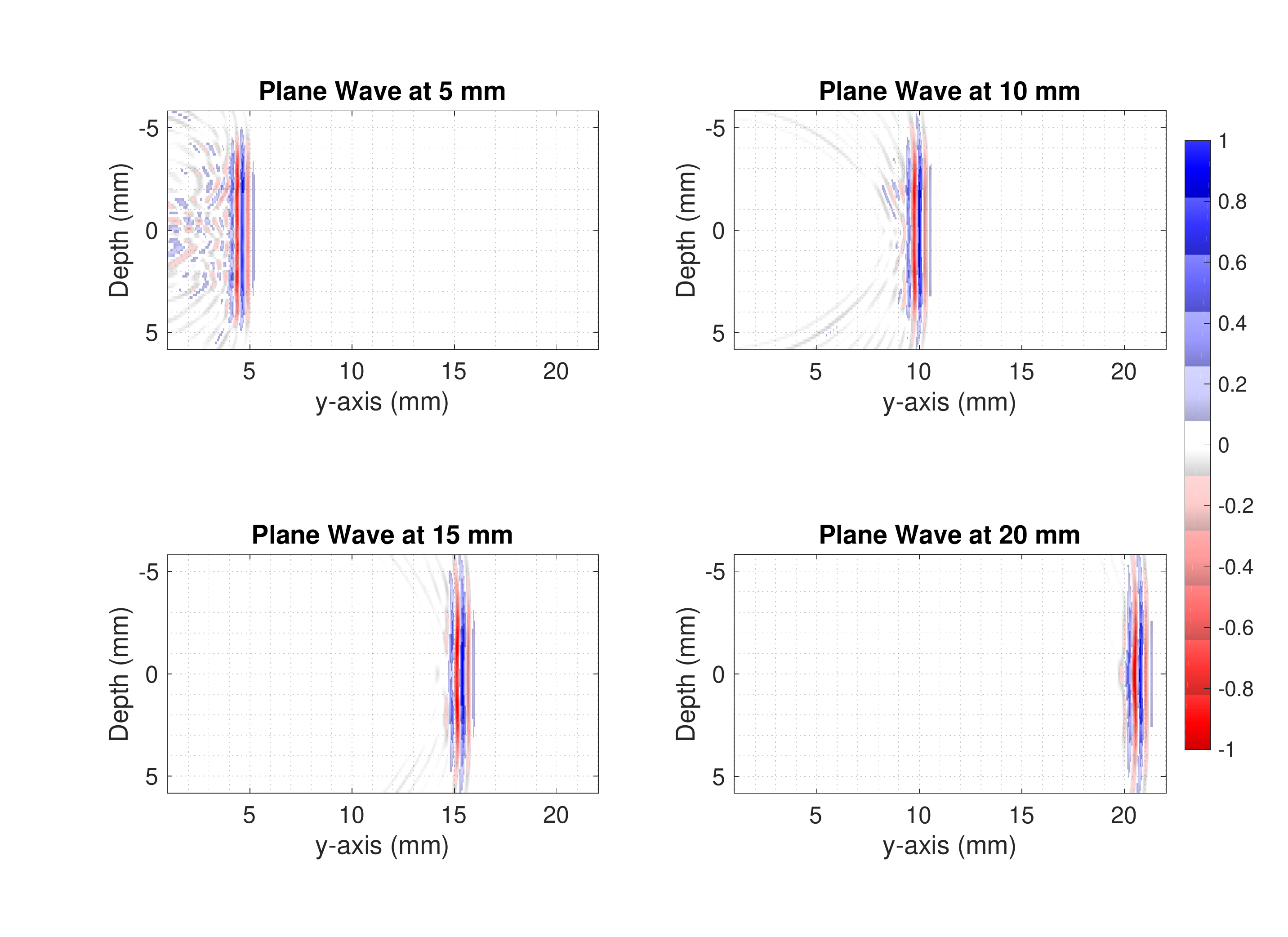}
	\caption{Simulated plane wave propagation at 5, 10, 15 and 20~mm depths. A 3-cycle Gaussian pulse was simultaneously transmitted from 512 elements of the 2D array. All the panels are normalized to their respective maximum.}
	\label{fig:SparseArray_Tukey_Plane_wave_propogation}
\end{figure}

\begin{figure}[!t]
	\centering
	\includegraphics[viewport = 80 100 950 450,  width = 80mm, clip]{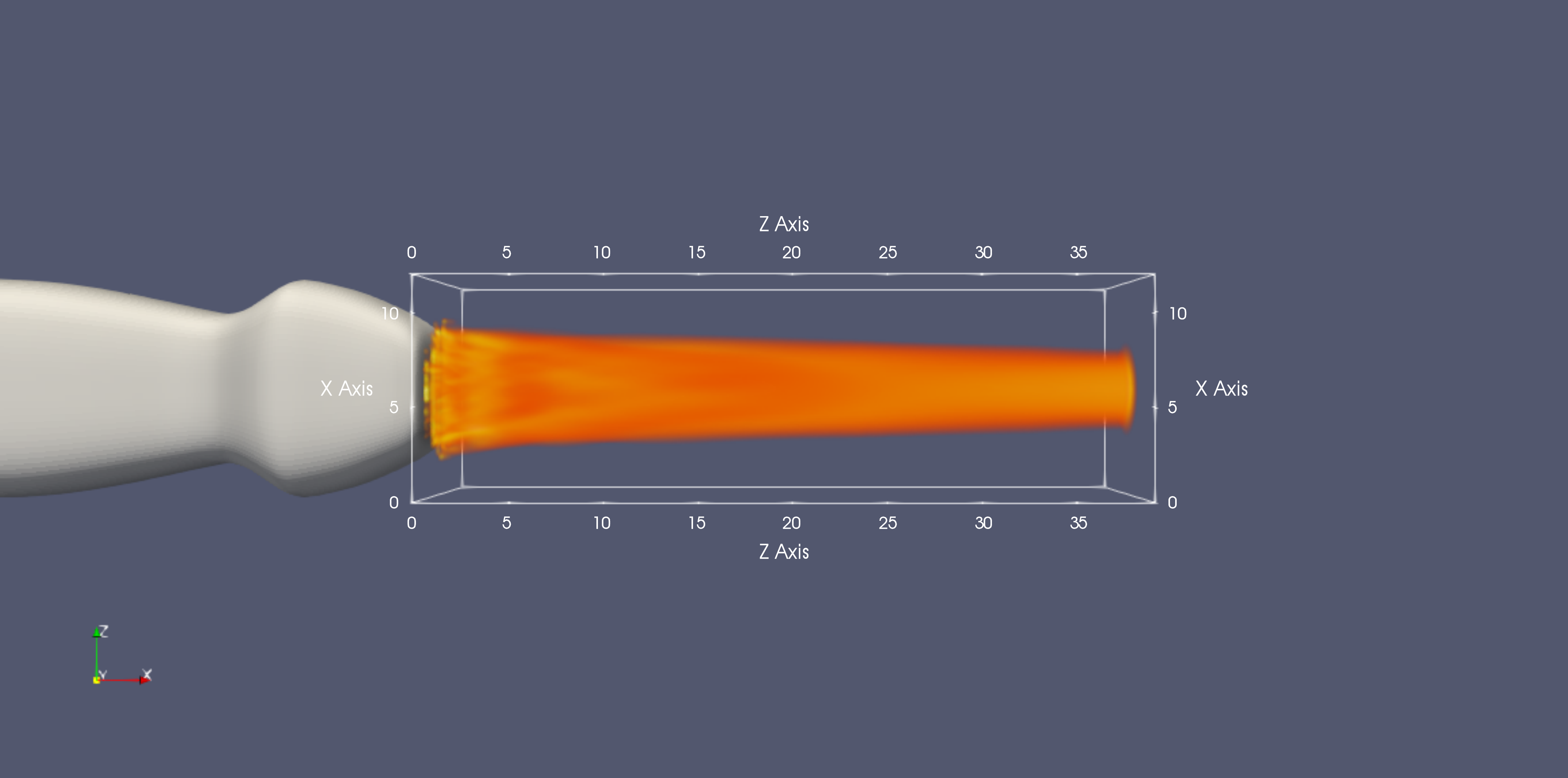}
	\includegraphics[viewport = 80 100 950 450,  width = 80mm, clip]{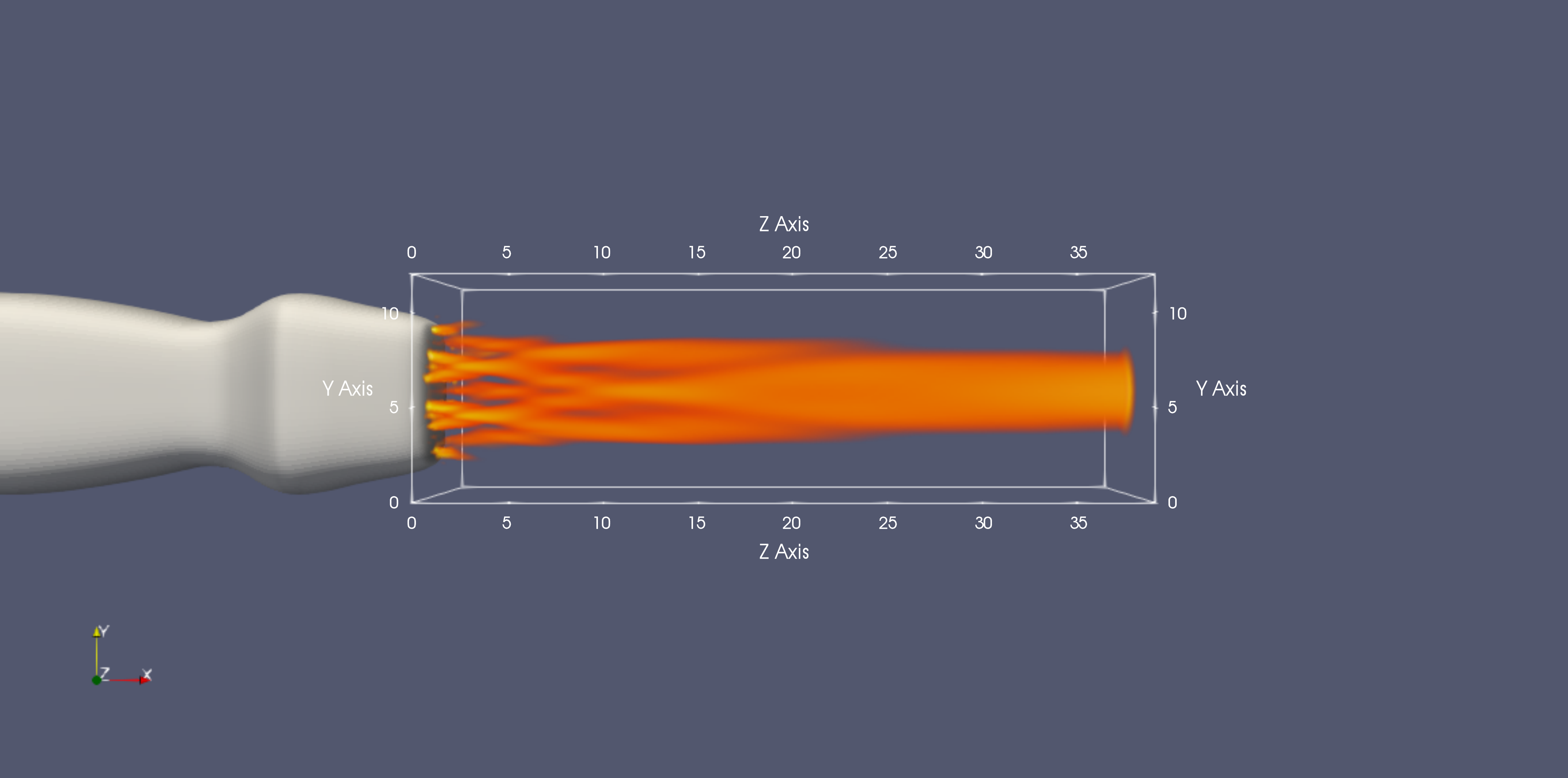}
	\caption{Simulated 3-D ultrasound field radiated from the sparse array is shown from (Top) the x-z view, and (Bottom) the y-z view, where z-axis represents depth.}
	\label{fig:SparseArray_3DField}
\end{figure}

To evaluate the feasibility of the proposed approach, plane wave propagation from the 512 element sparse array was simulated at different depths as shown in Fig.~\ref{fig:SparseArray_Tukey_Plane_wave_propogation} using Field II~\cite{Jensen1992,Jensen1996}. The radiated ultrasound field within the first 5~mm depth (Fig.~\ref{fig:SparseArray_Tukey_Plane_wave_propogation} (top-left)) is a combination of a plane wave and a dispersed tail, which is a result of missing rows. At the depth of 10~mm, as shown in Fig.~\ref{fig:SparseArray_Tukey_Plane_wave_propogation} (top-right), the tail resembles a superposition of multiple edge waves as a result of discontinuities in the array. At this point, the radiated beam shape is not suitable for generating a good quality image. Around 15~mm depth, as shown in Fig.~\ref{fig:SparseArray_Tukey_Plane_wave_propogation} (bottom-left), the tail becomes less prominent and edge waves diminish below $-14$~dB; however, it can still produce image artefacts as demonstrated by~\cite{Rasmussen2015}. Further away from the transducer, the residual waves behind the wavefront disappear and the ultrasound field becomes more uniform, which is suitable for plane wave imaging after 20~mm depth as shown in Fig.~\ref{fig:SparseArray_Tukey_Plane_wave_propogation} (bottom-right). The 3-D simulations displayed in Fig.~\ref{fig:SparseArray_3DField} also support the same conclusion: due to the choice of elements and three unconnected rows, the ultrasound field is not uniform for the first 20~mm.

	\subsection{3-D Super-resolution Experimental Results}

	
\begin{figure}[!t]
		\centering
		\includegraphics[viewport = 150 35 700 624,  width = 78mm, clip]{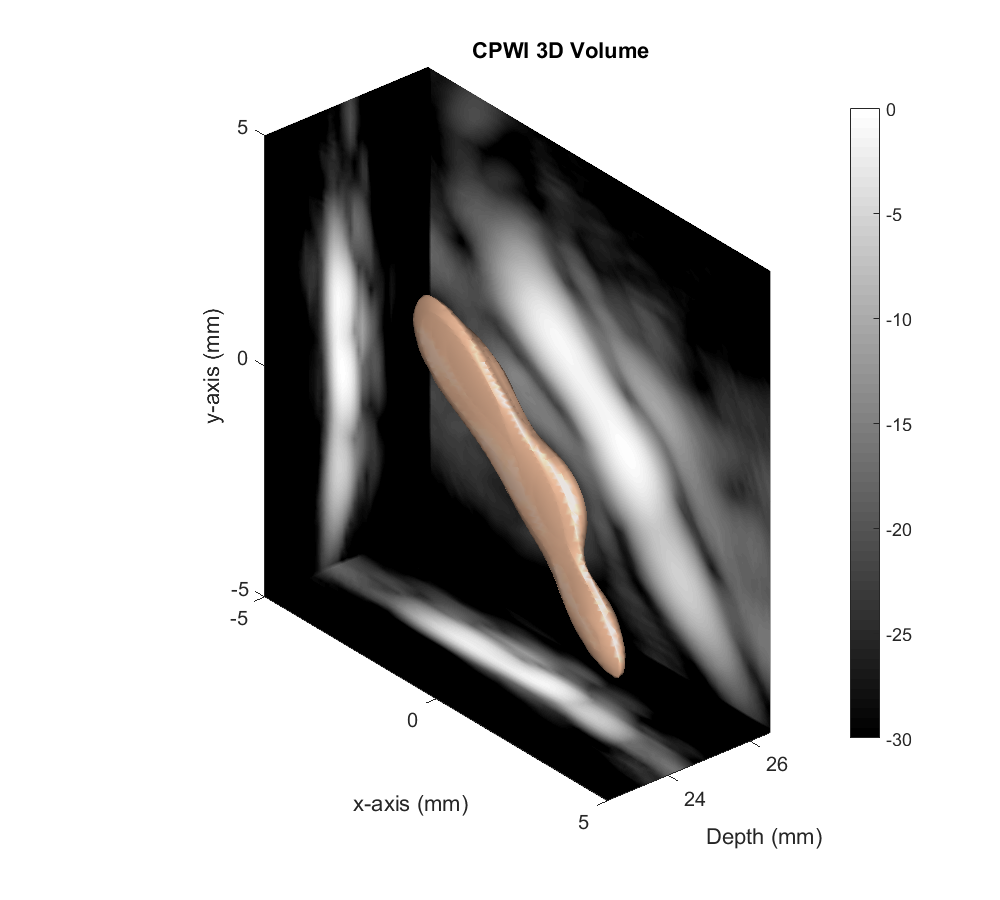}
		\caption{Isosurface of the 3-D ultrasound image is plotted in copper at -10 dB level. 2-D maximum intensity projections with a 30~dB dynamic range is overlaid on the volumetric image.}
		\label{fig:Bmode_3D}
\end{figure}	
	
\begin{figure}[!t]
	\centering
	\includegraphics[viewport = 140 20 700 600,  width = 88mm, clip]{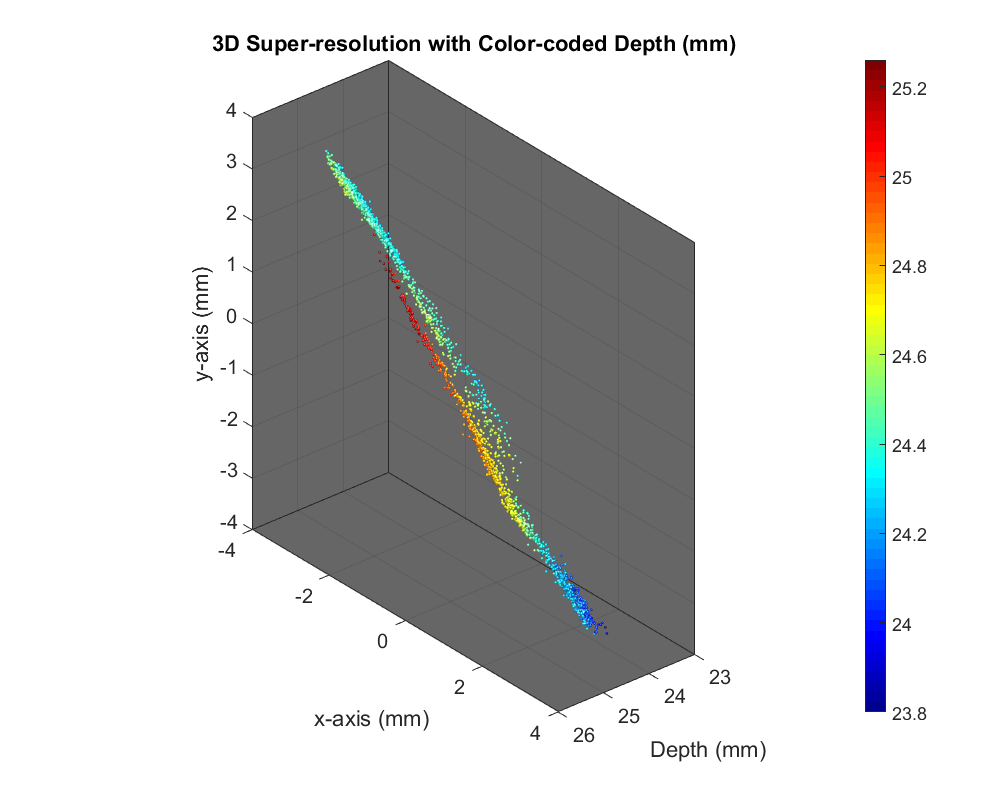}
	\includegraphics[viewport = 140 20 700 600,  width = 88mm, clip]{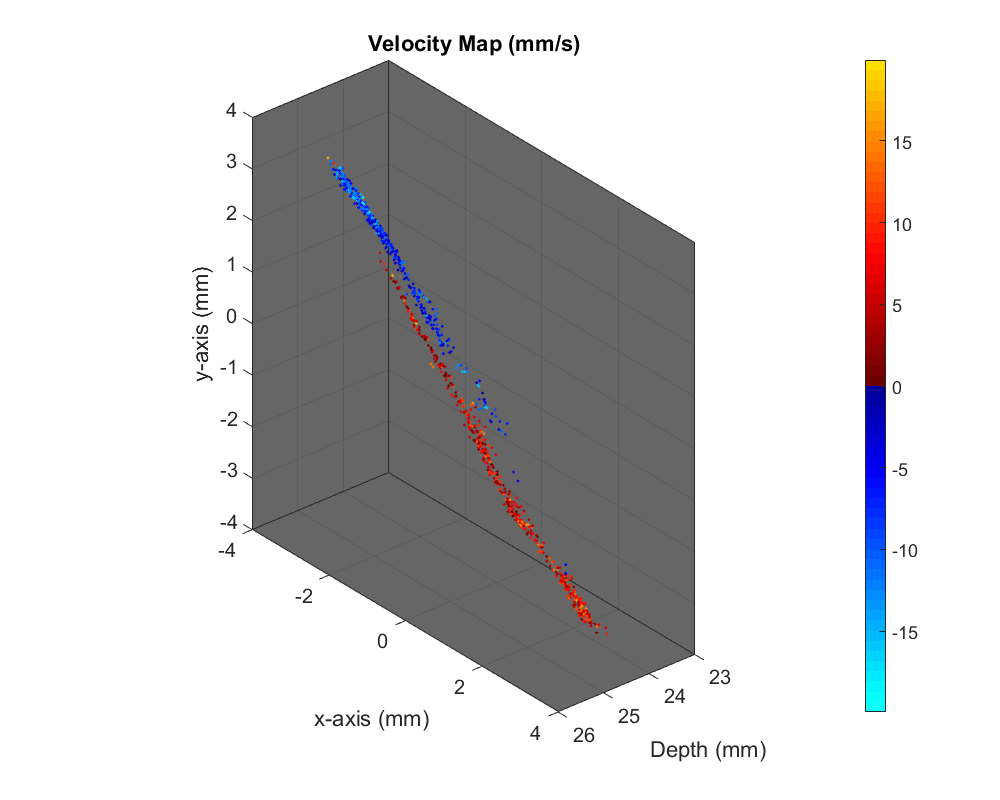}	\caption{(Top) 3-D super-resolution image of the two 200 $\mu$m tubes arranged in a double helix shape. Depth-encoded colorscale is added to improve the visualization. (Bottom) Velocity and direction (positive towards increasing y direction) of tracked microbubbles. }
	\label{fig:SR_3D_colorcoded_and_Velocity}
\end{figure}

Before performing the experiments on a cellulose microvasculature phantom, the imaging performance of the 2-D sparse array was characterized with a point target using the tip of a 100~$\mu$m metal wire. The full-width-half-maximum (FWHM) of the 3-D B-mode point-spread-function (PSF) was measured as 793, 772, and 499 $\mu$m in the x, y \& z directions respectively by using linear interpolation ~\cite{Harput2014a}. The localization precision was measured to be the standard deviation of the localization positions over 100 frames. The 3-D super-localization precision of the overall system at 25~mm was found to be 18~$\mu$m in the worst imaging plane (x direction), where the imaging wavelength is 404 $\mu$m in water at 25$^{\circ}$C.

The volumetric B-mode image of two cellulose tubes without microbubble flow is shown in Fig.~\ref{fig:Bmode_3D}. In addition to the 3-D visualization of the structure displayed in copper color, 2-D maximum-intensity-projection (MIP) slices in three directions were plotted. It was not possible to visualize the two separate 200 $\mu$m tubes in these MIP slices or in the volumetric image.

Fig.~\ref{fig:SR_3D_colorcoded_and_Velocity} (top) shows the 3-D super-resolved volume of the imaged sub-wavelength structures by combining localizations from all acquired frames. A total of 2824 microbubbles were localized within the 6000 volumes after compounding. Due to the large number of localizations, the 3-D structure of the tubes cannot be clearly visualized in a single 2-D image. To improve the visualization, 3-D SR-US images are plotted with depth information color-coded in the image.

Fig.~\ref{fig:SR_3D_colorcoded_and_Velocity} (bottom) displays the velocity profiles of tracked microbubbles. Only 1076 microbubble-pairs out of 2824 microbubbles were traceable from consecutive frames using a nearest-neighbor method. Using these microbubble tracks, two sub-wavelength tubes with opposing flows were easily distinguishable by color-coding the direction of their velocity vectors. The percentage of microbubbles that were followed over two or more volumes of 76.2\% was attributed to microbubble destruction, which can also be observed in the super-resolution images showing a high number of microbubble localizations at the inlet and almost no localizations at the outlet of the tubes, see Fig.~\ref{fig:SR_3D_colorcoded_and_Velocity}.

\begin{figure}[!t]
	\centering
	\includegraphics[viewport = 35 25 630 660,  width = 88mm, clip]{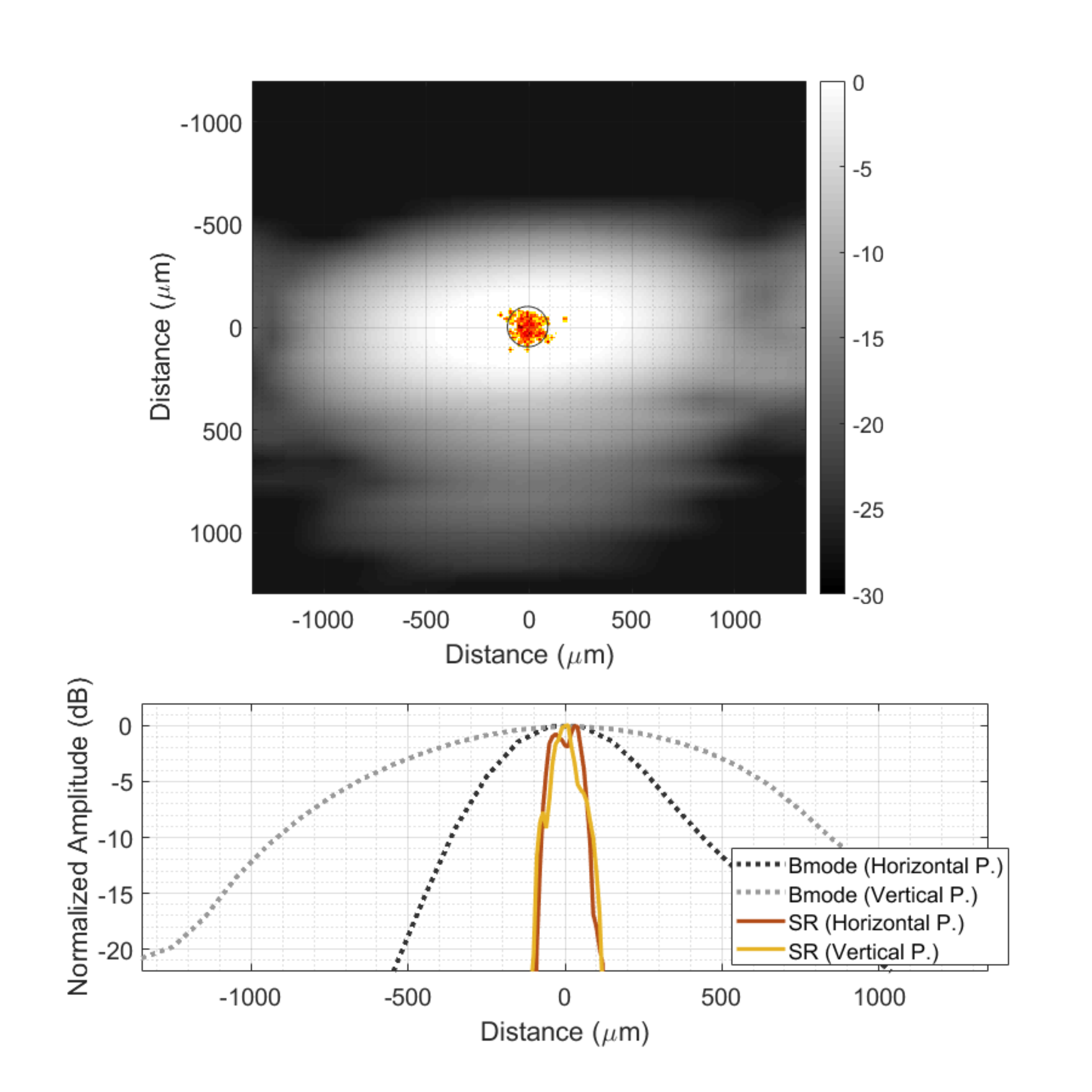}
	\caption{(Top) Figure shows the MIP of the B-mode image belonging to a 1~mm long section of the tube projected into a 2D plane that is orthogonal to the direction of the flow. The super-resolution image was projected into the same 2D plane and overlaid on the B-mode image. Black circle represents the 200~$\mu$m tube circumference. (Bottom) The B-mode FWHM of the tube is measured as 1380~$\mu$m and 590~$\mu$m from 1-D projections in the horizontal and vertical directions of the top panel plot respectively. The super-resolution FWHM of the tube is measured as 134~$\mu$m and 108~$\mu$m from 1-D projections in the horizontal and vertical directions of the top panel plot respectively.}
	\label{fig:Bmode_SR_Analysis_FWHM}
\end{figure}

\begin{figure}[!t]
	\centering
	\includegraphics[viewport = 35 25 630 660,  width = 88mm, clip]{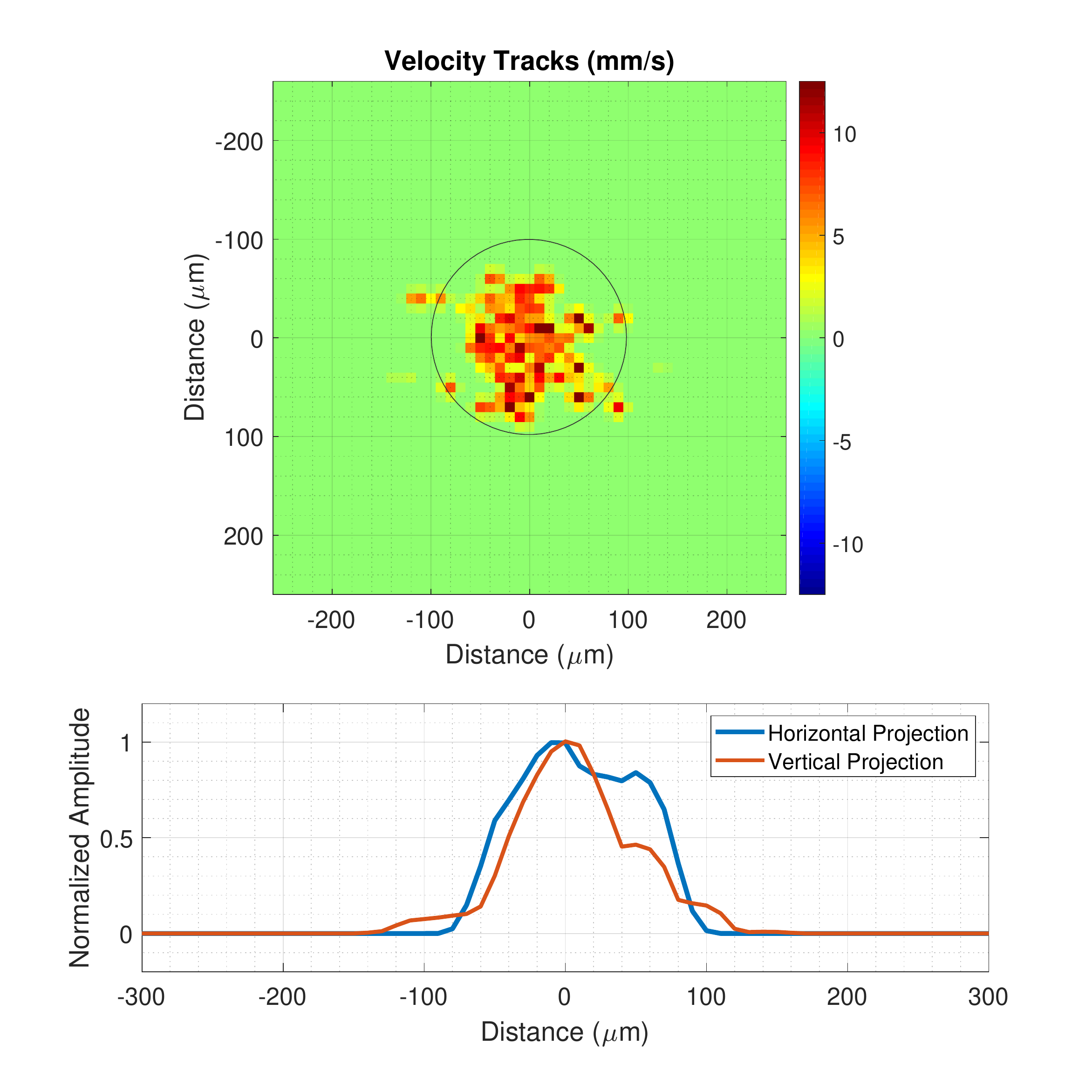}
	\caption{(Top) Figure shows the velocity tracks belonging to a 1~mm long section of the tube  projected into a 2D plane that is orthogonal to the direction of the flow. Black circle represents the 200~$\mu$m tube circumference. (Bottom) The FWHM of the tube is measured as 130~$\mu$m and 75~$\mu$m from 1-D projections in the horizontal and vertical directions of the top panel plot respectively.}
	\label{fig:3D_SR_Analysis_FWHM}
\end{figure}

The thickness of the imaged tubes was measured at the inlet where the tube is clearly isolated in the 3-D SR-US image around the [3~mm, -3~mm] coordinates in x and y respectively. To perform the thickness measurement, a 1~mm long section of the imaged tube was chosen and projected into a 2-D plane that is orthogonal to the direction of the tube as shown in Fig.~\ref{fig:Bmode_SR_Analysis_FWHM} (top) both for B-mode and 3-D SR-US images. Fig.~\ref{fig:Bmode_SR_Analysis_FWHM} (bottom) shows the 1-D projections in the horizontal and vertical directions where the FWHM of the super-resolved tube was measured as 134~$\mu$m and 108~$\mu$m and the $-20$~dB width of the super-resolved tube was measured as 213~$\mu$m and 200~$\mu$m respectively. In the B-mode image two touching tubes appeared as a single scattering object with a FWHM of 1380~$\mu$m and 590~$\mu$m in the horizontal and vertical 1-D projections respectively.

\begin{figure}[!t]
	\centering
	\includegraphics[viewport = 45 0 700 600,  width = 76mm, clip]{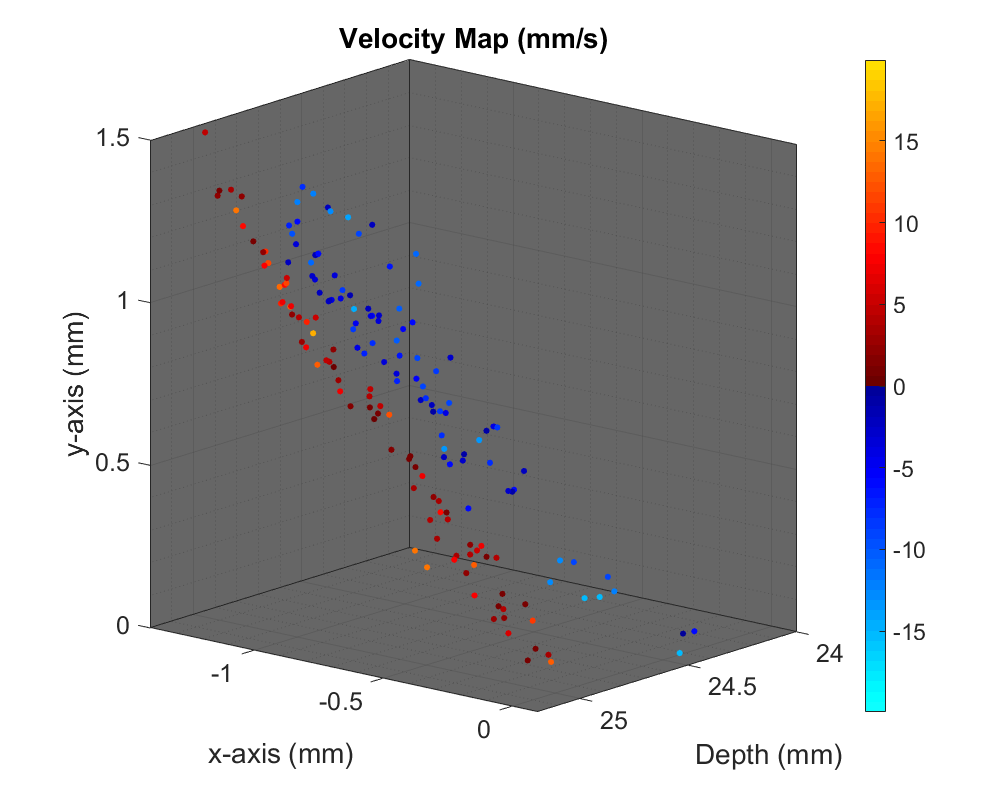}
	\includegraphics[viewport = 35 25 630 660,  width = 88mm, clip]{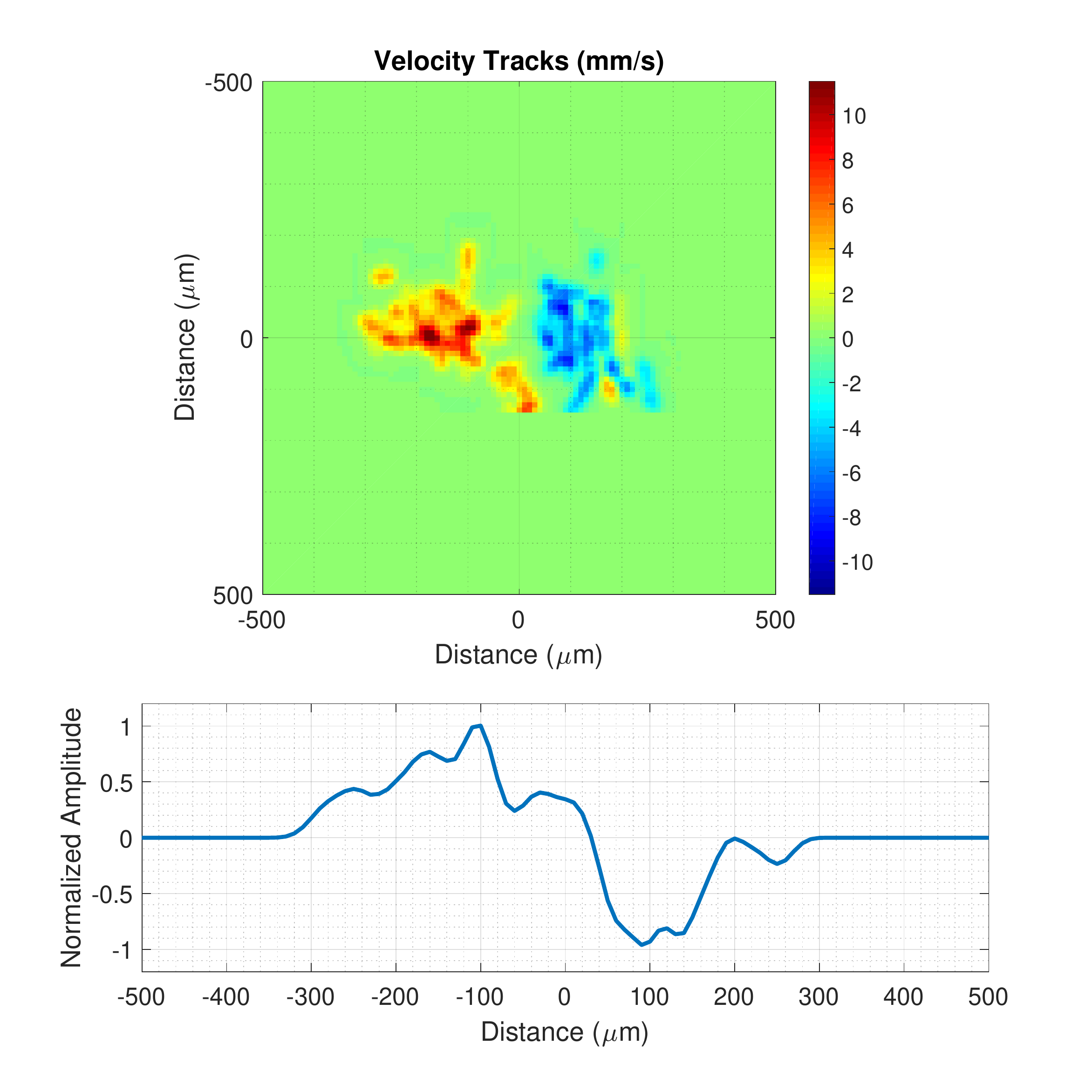}
	\caption{ (Top) 3D velocity profiles of microbubbles re-plotted from Fig.~\ref{fig:3D_SR_Analysis_FWHM} (bottom) to display the details more clearly when two tubes are in contact.	(Middle) Figure shows the velocity tracks belonging to a 1~mm long section of the tube  projected into a 2-D plane. (Bottom) The 1-D projection of the middle panel plot towards vertical direction shows that the peak-to-peak distance between two opposing velocity tracks is 190~$\mu$m.}
	\label{fig:3D_SR_Analysis}
\end{figure}

Microbubble tracking effectively worked as another layer of filtering by removing the potentially erroneous non-traceable super-localizations. Around the same section of the tube shown in Fig.~\ref{fig:Bmode_SR_Analysis_FWHM} (top) and Fig.~\ref{fig:3D_SR_Analysis_FWHM} (top), 96\% of the microbubble velocity tracks were within a diameter of 200 $\mu$m, which was 89\% for the super-localized microbubbles without velocity tracking. The FWHM of a single tube appeared as 130~$\mu$m and 75~$\mu$m from the projection of velocity tracks to horizontal and vertical directions, as plotted in Fig.~\ref{fig:3D_SR_Analysis_FWHM} (bottom).

Microbubble tracking made the separation between the tubes more clear when tubes are in contact around the central section of the 3-D SR-US and velocity maps displayed in Fig.~\ref{fig:SR_3D_colorcoded_and_Velocity}. The velocity profiles of microbubbles at this location with two touching tubes were re-plotted in Fig.~\ref{fig:3D_SR_Analysis} (top) for clear visualization. From this plotted volume, a 1~mm long section of the tube was chosen and projected into a 2-D plane that is shown in Fig.~\ref{fig:3D_SR_Analysis} (middle). In this 2-D maximum intensity projection, the weighted center locations between positive and negative velocity tracks have a distance of 239~$\mu$m. The 1-D projection of the velocity tracks had a FWHM of 122~$\mu$m and 115~$\mu$m for positive and negative flows respectively with a peak-to-peak distance of 190~$\mu$m between two opposing tracks as plotted in Fig.~\ref{fig:3D_SR_Analysis} (bottom).

	\section{Discussion}

A better 3-D image quality may be achieved by using a large number of independent array elements with the fastest possible volumetric imaging rate; however this requires the same number of hardware channels as the number of elements and the ability to process very large stacks of data. Due to the high cost, full 2-D array imaging using an ultrasound system to control very large numbers of independent elements has only been used by a few research groups~\cite{Jensen2013,Provost2014,Gennisson2015,Petrusca2018}. These systems had 1024 channels capable of driving a $32 \times 32$ 2-D array with at least 4 connectors. Even some of these systems had 1 of 2 transducer elements multiplexed in reception~\cite{Provost2014,Gennisson2015}. Many researchers have developed methods to use a large number of active elements with fewer channels (usually between 128 and 256) to reduce the cost and complexity of the ultrasound systems and the probes. It has been demonstrated in several studies that row-column addressed matrix arrays~\cite{Savoia2007,Rasmussen2015,Holbek2016,Flesch2017}, microbeamformers~\cite{Savord2000,Matrone2014,Santos2016} and channel multiplexing can be an alternative to fully addressed 2-D matrix arrays. However, these methods have less flexibility and limitations due to the elements not being continuously connected to the ultrasound system.

In this paper, a 2-D sparse array imaging probe has been developed for 3-D super-resolution imaging. This has addressed the main limitation of the existing 2-D imaging of poor spatial resolution in the elevational plane. In addition to super-resolution imaging, 3-D velocity mapping was implemented to reveal the flow inside the microstructures. Using the sparse array approach instead of a full matrix array reduced the number of channels to half, and hence the connection issues, cost and data size while still achieving the same volumetric acquisition speed since all elements of 2-D spiral array are always connected to the system. Although this can sacrifice the maximum achievable transmit pressure and receive sensitivity, it is not a significant issue with SR-US due to the low pressure required and high sensitivity achievable in microbubble imaging. In terms of B-mode image resolution, the axial resolution is comparable, since both arrays have the same bandwidth; while a slightly worse lateral resolution is expected for the sparse array, since the full matrix array has a larger aperture size. It is hard to distinguish the grating lobes and the side lobes of a sparse array, but here we consider the unwanted leakage outside the main lobe as grating lobes if it is as a result of element-to-element spacing, and as side lobes if it is as a result of finite aperture size. The side lobe and edge wave suppression characteristics of the sparse array will outperform an un-apodized full matrix array thanks to the integrated apodization, although the fixed apodization might be a limitation for some applications. Both arrays will have higher grating lobes in y direction due to the three empty rows. The highest grating lobe of the full matrix array will appear at $\pm 8^{\circ}$ as high as 17\% of the main lobe amplitude, calculated using the array factor equation in \cite{Harput2008}. A sparse choice of elements spreads the grating lobes to a wider range due to the irregular placement of elements, where the highest grating lobe will appear at $\pm 18^{\circ}$ as high as 16\% of the main lobe amplitude.

Using the plane-wave imaging method instead of line-by-line scanning increases the temporal resolution of the volumetric imaging. Faster 3-D image acquisition provides a higher microbubble localization rate and improves velocity estimations due to more frequent sampling. However, a microbubble travelling with a velocity of 10 mm/s will be exposed to 3000 ultrasound pulses while travelling through the imaging region of 10~mm at a PRF of 3000 Hz. At this insonation rate, even at a relatively low MI of 0.07 almost all microbubbles were destroyed before reaching the center of the imaging region. At this point, a new transmitting strategy was implemented to reduce the microbubble destruction rate instead of reducing the PRF, which may introduce artefacts on 9 compounded volumes due to moving microbubbles, or ultrasound pressure, which will decrease the SNR and localization precision. To improve the microbubble circulation time, a 7~millisecond pause was added between each compounded volume acquisition that took 3~milliseconds. This strategy reduced insonation rate by 3.3-fold and the volumetric acquisition rate to 100 Hz without sacrificing the PRF during compounding. The second advantage of this transmit strategy is a reduction in the redundant microbubble localizations at the same spatial location without reducing the PRF. In order for microbubbles to provide new spatial information in each frame, they must be re-located by the flow. Especially for microvasculature with slow physiological flow, increasing the frame rate will no longer increase the obtained spatial information but result in redundant location information. Nevertheless, using high frame rates may still be valuable for improving the SNR and velocity tracking. The proposed transmit strategy improved the image quality for this study, but it could not totally prevent the microbubble destruction that caused a lower number of localizations at outlet of the tubes visible in the 3-D SR-US image. In the future, the relationship between PRF, pause interval, microbubble flow velocity, and compounding strategies should be investigated for different applications and physiological flow rates.

	\section{Conclusion}
	
The main limitation of localization-based SR-US imaging performed in 2-D is the lack of super-resolution in the elevation direction. In this study, this issue was addressed by using a bespoke 2-D sparse array that achieved an estimated localization precision of 18~$\mu$m in the worst imaging plane, which is approximately 22 times smaller than the wavelength. Compounded plane wave imaging with a PRF of 3000 Hz enabled super-resolution imaging in all spatial directions with an image acquisition time of 60 seconds. The structure of two 200 $\mu$m, smaller than half wavelength, tubes arranged in a double helix shape were super resolved and flow velocities within these tubes were estimated. 3-D sub-diffraction imaging was achieved \textit{in vitro} using the 2D sparse array probe.


	\section*{Acknowledgments}  
	   
This work was supported mainly by the EPSRC under Grant EP/N015487/1 and EP/N014855/1, in part by the King's College London (KCL) and Imperial College London EPSRC Centre for Doctoral Training in Medical Imaging (EP/L015226/1), in part by the Wellcome EPSRC Centre for Medical Engineering at KCL (WT 203148/Z/16/Z), in part by the Department of Health through the National Institute for Health Research comprehensive Biomedical Research Center Award to Guy's and St Thomas' NHS Foundation Trust in partnership with KCL and King's College Hospital NHS Foundation Trust, in part by the Graham-Dixon Foundation and in part by NVIDIA GPU grant.

	
	\bibliography{BuBBle,Ultrasound,SignalProcessing,MotionCorrection,SuperRes,3D_Imaging,Beamforming}       
	\bibliographystyle{IEEEtran}

\end{document}